\documentclass[10pt,twocolumn,letterpaper]{article}

\usepackage[pagenumbers]{cvpr} % To force page numbers, e.g. for an arXiv version

\usepackage{bbm}

\definecolor{cvprblue}{rgb}{0.21,0.49,0.74}
\usepackage[pagebackref,breaklinks,colorlinks,allcolors=cvprblue]{hyperref}

\def\paperID{*****} % *** Enter the Paper ID here
\def\confName{CVPR}
\def\confYear{2026}

\title{Vision-Language Model Based Multi-Expert Fusion for CT Image Classification}

\author{Jianfa Bai$^1$, Kejin Lu$^1$, Runtian Yuan$^1$, Qingqiu Li$^1$, Jilan Xu$^2$, Junlin Hou$^3*$, Yuejie Zhang$^1*$, Rui Feng$^1*$\\
$^1$College of Computer Science and Artificial Intelligence,\\
Shanghai Key Laboratory of Intelligent Information Processing, Fudan University\\
$^2$University of Oxford\\
$^3$The Hong Kong University of Science and Technology
}

\begin{document}
\maketitle
\begin{abstract}
Robust detection of COVID-19 from chest CT remains challenging in multi-institutional settings due to substantial source shift, source imbalance, and hidden test-source identities. In this work, we propose a three-stage source-aware multi-expert framework for multi-source COVID-19 CT classification. First, we build a lung-aware 3D expert by combining original CT volumes and lung-extracted CT volumes for volumetric classification. Second, we develop two MedSigLIP-based experts: a slice-wise representation and probability learning module, and a Transformer-based inter-slice context modeling module for capturing cross-slice dependency. Third, we train a source classifier to predict the latent source identity of each test scan. By leveraging the predicted source information, we perform model fusion and voting based on different experts. On the validation set covering all four sources, the Stage 1 model achieves the best macro-F1 of 0.9711, ACC of 0.9712, and AUC of 0.9791. Stage~2a and Stage~2b achieve the best AUC scores of 0.9864 and 0.9854, respectively. Stage~3 source classifier reaches 0.9107 ACC and 0.9114 F1. These results demonstrate that source-aware expert modeling and hierarchical voting provide an effective solution for robust COVID-19 CT classification under heterogeneous multi-source conditions.
\end{abstract}    
\section{Introduction}
\label{sec:intro}

Robust detection of COVID-19 from chest CT is not merely a classification problem, but a domain-generalization problem under substantial variation between institutions \cite{kollias2024domain,kollias2025pharos,arsenos2022large,arsenos2023data,gerogiannis2024covid,kollias2021mia,kollias2022ai,kollias2023deep,kollias2023ai,kollias2024sam2clip2sam,yuan2025multi,wang2022medclip,gunraj2022covidnetct2,krueger2021vrex,zhang2017mixup,cao2021exploiting,kollias2020deep,kollias2020transparent,kollias2018deep,li2024advancing,yuan2024domain,li2025advancing}. In multi-center clinical settings, CT scans may differ markedly in scanner vendors, reconstruction kernels, slice counts, field-of-view, and background artifacts, while patient populations and disease presentations also vary across hospitals. As a result, a model trained on one source can perform well in-distribution yet degrade noticeably on unseen or shifted sources. This issue becomes even more critical in the Multi-Source COVID-19 Detection Challenge, where the task explicitly involves four distinct medical sources and therefore directly evaluates whether a method can maintain reliable performance under source shift rather than simply fit a single pooled dataset.

%This benchmark is challenging not only because of inter-source heterogeneity, but also because of several nontrivial data issues that complicate model development. In our setting, the official validation split of source 2 contains no COVID-positive cases, which makes source-aware validation and model selection particularly unstable for that domain. 
% Moreover, the source 0 non-COVID training set requires folder-level correction, and the test set does not provide source identities in advance, making source-specific inference impossible without an additional source prediction stage. These properties make the task fundamentally different from standard COVID-19 CT classification benchmarks and motivate a framework that is robust to domain shift, tolerant to source imbalance, and capable of source-aware expert allocation.

\begin{figure*}[t]
    \centering
    \includegraphics[width=\textwidth]{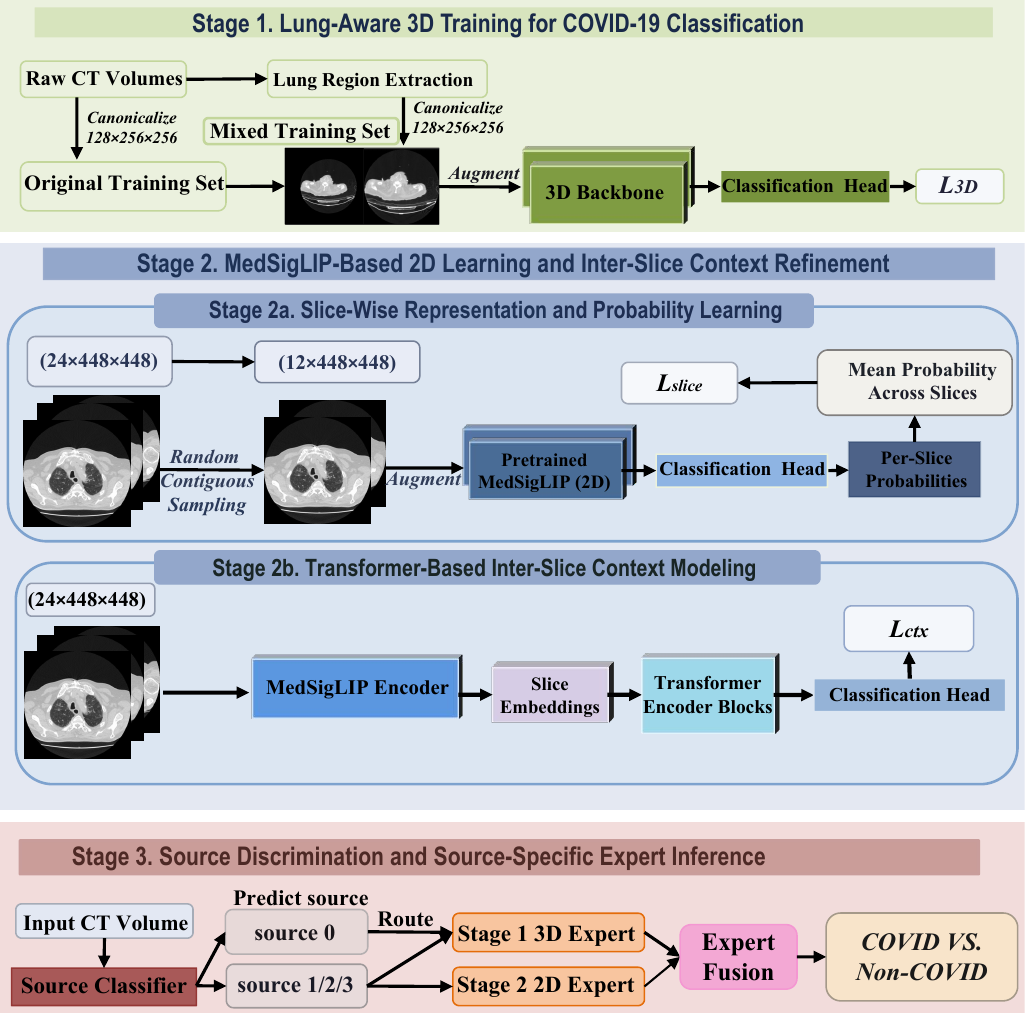}
    \caption{Overview of the proposed three-stage source-aware multi-expert framework for multi-source COVID-19 CT classification. Stage 1 builds a lung-aware 3D expert for volumetric classification. Stage 2 introduces two MedSigLIP-based 2D experts for slice-wise probability learning and inter-slice context modeling. Stage 3 performs source discrimination and source-specific expert inference, where source 0 is handled by the 3D expert and source 1/2/3 are jointly inferred by multiple experts through fusion and voting.}
    \label{fig:architecture}
\end{figure*}

% Existing CT-based COVID-19 studies have demonstrated that deep models can achieve strong diagnostic performance on curated datasets. For example, Li \textit{et al.} developed a deep learning system to distinguish COVID-19 from community-acquired pneumonia and other lung conditions on chest CT \cite{li2020ai}. Bai \textit{et al.} showed that AI assistance can improve radiologists' differentiation of COVID-19 from other pneumonias on CT \cite{bai2020ai}. Harmon \textit{et al.} further validated AI-based COVID-19 CT detection on multinational datasets, highlighting the importance of cross-site evaluation \cite{harmon2020ai}. More recent systems such as COVID-Net CT-2 improved data diversity and network design for CT-based COVID-19 screening \cite{gunraj2022covidnetct2}. However, most prior methods still emphasize direct end-to-end classification and do not explicitly address hidden test sources, source-specific model preference, or validation sparsity under highly imbalanced source partitions.

To address these limitations, we propose a \textbf{three-stage source-aware multi-expert framework} for multi-source COVID-19 CT classification. As illustrated in Fig.~\ref{fig:architecture}, Stage 1 builds a lung-aware 3D expert that performs lung extraction and unified volume canonicalization before 3D classification, reducing irrelevant peripheral background and preserving volumetric morphology. Stage 2 introduces two complementary 2D experts based on a pretrained MedSigLIP \cite{sellergren2025medgemma} encoder: Stage 2a performs slice-wise probability learning by randomly sampling contiguous slices during training and averaging per-slice predictions, while Stage 2b removes the original classification head and adds lightweight Transformer \cite{vaswani2017attention} encoder blocks to refine inter-slice context before classification. Stage 3 further trains a source classifier to estimate the latent source identity of each test scan and route samples to appropriate experts.
%In particular, source 0 is handled by the Stage 1 3D expert alone, whereas source 1/2/3 are jointly inferred by Stage 1, Stage 2a, and Stage 2b through expert voting and fusion. 
To improve stability, we also train multiple model variants within each step and aggregate their outputs by voting.

Our framework is designed together with challenge-specific data handling. We first perform lung extraction for all scans and suppress excessive peripheral dark regions. We then apply a unified preprocessing rule that discards the first and last 15\% of slices when the slice count exceeds 150, followed by stage-specific canonicalization for 3D and 2D pipelines. To alleviate the zero-positive validation issue of source 2, we include its 39 positive training cases in validation-oriented source analysis. We also correct the source 0 non-COVID training count based on folder-level inspection and predict test-source identities for source-aware inference. These steps are important for reliable model development under scarce and imperfect source supervision.

Extensive experiments validate the effectiveness of the proposed design. Stage 1 achieves a best macro-F1 of \textbf{97.11\%}, AUC of \textbf{98.29\%}, and accuracy of \textbf{97.12\%}. Stage 2a reaches a best macro-F1 of \textbf{94.50\%}, AUC of \textbf{98.64\%}, and accuracy of \textbf{94.81\%}, while Stage 2b further improves to \textbf{95.82\%} macro-F1, \textbf{98.54\%} AUC, and \textbf{95.97\%} accuracy. The Stage 3 source classifier achieves \textbf{91.07\%} accuracy and \textbf{91.14\%} F1, providing a reliable basis for source-aware routing. Notably, the Stage 1 3D expert attains perfect accuracy on source 0, which directly motivates our final source-specific expert allocation strategy.

Our main contributions are summarized as follows:
\begin{itemize}
    \item We reformulate multi-source COVID-19 CT detection as a \textbf{source-aware domain-robust classification problem} and explicitly address hidden test sources, source imbalance, and validation sparsity.
    \item We propose a \textbf{three-stage multi-expert framework} that combines a lung-aware 3D expert, two MedSigLIP-based 2D experts, and a source classifier for source-specific inference.
    \item We introduce a \textbf{hierarchical voting and fusion strategy} that integrates multiple models within each step and jointly exploits heterogeneous experts for source 1/2/3 prediction.
    \item We achieve strong validation performance across stages, with the best Stage 1 model reaching \textbf{97.10\%} macro-F1, demonstrating the effectiveness of source-aware expert modeling for robust multi-source COVID-19 CT classification.
\end{itemize}

\section{Methodology}
\label{sec:3_method}
\subsection{Overview}

As illustrated in Fig.~\ref{fig:architecture}, we propose a three-stage source-aware multi-expert framework for robust multi-source COVID-19 CT classification. Given a chest CT scan $X$ with binary label $y \in \{0,1\}$, where $y=1$ denotes \textit{COVID} and $y=0$ denotes \textit{Non-COVID}, our goal is to learn complementary experts under heterogeneous source distributions and perform source-aware inference at test time.

All scans are first processed by lung region extraction to suppress excessive irrelevant dark background around the body region. In addition, when the slice number exceeds 150, we discard the first 15\% and the last 15\% of slices to reduce unstable peripheral slices and excessive redundancy. After preprocessing, two stage-specific canonicalization strategies are adopted. For the 3D branch, each scan is converted to a volume of size $128 \times 256 \times 256$. For the 2D branch, each scan is canonicalized to $24 \times 448 \times 448$.

Our framework contains four trainable models. The first is a lung-aware 3D classifier in Stage~1, which directly predicts COVID/Non-COVID from canonicalized 3D CT volumes. The second is the Stage~2a model, which fine-tunes a pretrained MedSigLIP encoder by slice-wise representation and probability learning. The third is the Stage~2b model, which reuses the MedSigLIP encoder from Stage~2a, discards its original classification head, and introduces two Transformer encoder blocks for inter-slice context modeling. The fourth is a source classifier in Stage~3, which predicts the source identity of each scan and enables source-aware expert routing.

At inference time, the final prediction is determined according to the predicted source. Since the Stage~1 3D expert achieves perfect accuracy on source 0 in our validation analysis, scans predicted as source 0 are classified by the Stage~1 expert alone. For scans predicted as source 1/2/3, the predictions from Stage~1 and Stage~2 are combined by expert voting and fusion. Moreover, multiple model variants are trained within each stage, and their predictions are aggregated by voting to further improve robustness.

\subsection{Stage 1: Lung-Aware 3D Training for COVID-19 Classification}

Stage~1 is designed to learn a robust volumetric expert from both raw CT scans and lung-extracted CT scans, following the line of research established by prior CMC-based methods \cite{hou2021cmc,hou2021periphery,hou2022cmc_v2,hou2022boosting}. After preprocessing and canonicalization, each scan is represented as a 3D volume $V \in \mathbb{R}^{128 \times 256 \times 256}$. To improve robustness against source-dependent background variation, we construct a mixed training set that contains both canonicalized original volumes and canonicalized lung-extracted volumes.

Given an input volume $V$, a 3D ResNet-style backbone \cite{vaswani2017attention} extracts a volumetric representation, which is then passed to a binary classification head for COVID/Non-COVID prediction. The output logit is denoted by $\mathbf{z}_{3D} \in \mathbb{R}^{2}$, and the corresponding probability is computed by softmax. The Stage~1 objective is standard cross-entropy:
\begin{equation}
\mathcal{L}_{3D}
=
-\sum_{c=0}^{1}\mathbbm{1}(y=c)\log p_{3D}(y=c \mid V).
\end{equation}

During training, scan-level augmentation is applied to the 3D volumes, including cropping, resizing, and random rotation. Importantly, the same augmentation parameters are consistently applied to all slices within the same CT scan, while different scans use different random parameters. During validation, only deterministic resizing/rescaling is used. The optimized Stage~1 model serves not only as a strong 3D COVID-19 expert, but also as the feature backbone of the source classifier in Stage~3.

\subsection{Stage 2a: Slice-Wise Representation and Probability Learning}

Stage~2a adapts a pretrained MedSigLIP encoder to the COVID-19 CT task in a memory-efficient slice-wise manner. Each scan is first canonicalized to $24 \times 448 \times 448$. During training, instead of using all 24 slices at once, we randomly sample a contiguous 12-slice subsequence from the scan, denoted by $\tilde{U}=\{u_{\tau},u_{\tau+1},\dots,u_{\tau+11}\}$, where each slice $u_t \in \mathbb{R}^{448 \times 448}$.

Each sampled slice is independently fed into the pretrained MedSigLIP image encoder, followed by a binary classification head to obtain a slice-level probability. These 12 slice-level probabilities are then averaged to produce a scan-level prediction. Formally, if $p_t(y=c \mid u_t)$ denotes the predicted probability of class $c$ for the $t$-th sampled slice, then the scan-level probability is
\begin{equation}
\bar{p}_{\text{slice}}(y=c \mid \tilde{U})
=
\frac{1}{12}\sum_{t=1}^{12} p_t(y=c \mid u_t).
\end{equation}
The corresponding training loss is
\begin{equation}
\mathcal{L}_{\text{slice}}
=
-\sum_{c=0}^{1}\mathbbm{1}(y=c)\log \bar{p}_{\text{slice}}(y=c \mid \tilde{U}).
\end{equation}

This design allows Stage~2a to adapt the pretrained MedSigLIP encoder using scan-level supervision while avoiding excessive GPU memory consumption. At inference time, all 24 slices are used, and their probabilities are averaged in the same way to obtain the final Stage~2a prediction.

\subsection{Stage 2b: Transformer-Based Inter-Slice Context Modeling}

Although Stage~2a provides strong slice-level supervision, it does not explicitly model contextual dependency across adjacent slices. Therefore, in Stage~2b, we further build a sequence-level expert on top of the Stage~2a model.

Specifically, the MedSigLIP encoder trained in Stage~2a is reused as the visual backbone. Its original classification head is discarded, and most encoder layers are frozen, with only the last two layers kept trainable. Each of the 24 canonicalized slices is passed through this partially frozen encoder to obtain a sequence of slice embeddings. These embeddings are then fed into two Transformer encoder blocks to model inter-slice contextual relationships. The resulting contextualized slice features are aggregated into a scan-level representation, which is finally fed to a new binary classification head for COVID/Non-COVID prediction.

Let $U$ denote the full 24-slice scan and let $p_{\text{ctx}}(y=c \mid U)$ denote the final scan-level probability predicted by this module. The Stage~2b objective is
\begin{equation}
\mathcal{L}_{\text{ctx}}
=
-\sum_{c=0}^{1}\mathbbm{1}(y=c)\log p_{\text{ctx}}(y=c \mid U).
\end{equation}

Only the last two unfrozen MedSigLIP layers, the two Transformer encoder blocks, and the newly added classification head are updated in this stage. In this way, Stage~2b preserves the strong pretrained slice representation from Stage~2a while further introducing explicit cross-slice context modeling.

\subsection{Stage 3: Source Discrimination and Source-Specific Expert Inference}

The purpose of Stage~3 is to predict the hidden source identity of each scan and enable source-aware inference. To this end, we reuse the 3D backbone trained in Stage~1 and replace its binary classification head with a new four-class source classification head. The Stage 1 3D backbone is kept fixed and only the source classification head is trained. The resulting model predicts the source label $s \in \{0,1,2,3\}$ for each scan.

% Let $p_{\text{src}}(s=j \mid V)$ denote the predicted probability that a scan belongs to source $j$. The source classification objective is defined as
% \begin{equation}
% \mathcal{L}_{\text{src}}
% =
% -\sum_{j=0}^{3}\mathbbm{1}(s=j)\log p_{\text{src}}(s=j \mid V).
% \end{equation}

Because the official validation split of source 2 contains no positive scans, we additionally include the 39 positive training scans from source 2 in source-aware validation analysis and model selection. After training, the source classifier is applied to the test set to estimate source identities.

At test time, source-aware expert inference is performed according to the predicted source. If a scan is predicted as source 0, the final label is directly determined by the Stage~1 3D expert, since this expert achieves perfect accuracy on source 0 in our validation experiments. If a scan is predicted as source 1/2/3, the predictions from Stage~1, Stage~2a, and Stage~2b are integrated by expert voting. Denoting the corresponding step-level voted predictions by $\tilde{y}^{(1)}$, $\tilde{y}^{(2a)}$, and $\tilde{y}^{(2b)}$, the final prediction can be written as
\begin{equation}
\hat{y}
=
\operatorname{Vote}\big(\tilde{y}^{(1)}, \tilde{y}^{(2a)}, \tilde{y}^{(2b)}\big).
% \qquad \text{for scans predicted as source 1/2/3}.
\end{equation}

In practice, each stage contains multiple model variants with slightly different data processing settings but comparable performance, and their outputs are first aggregated within each step before cross-expert fusion. This hierarchical voting strategy improves robustness and helps stabilize prediction under source heterogeneity and source-specific data imbalance.
\section{Datasets and Experiments}
\label{sec:4_datasetsandexperiments}
\subsection{Datasets}

The Multi-Source COVID-19 Detection Challenge provides a multi-institutional chest CT dataset collected from four different sources. The official split is summarized in Table~\ref{tab:official_dataset}. The training and validation sets contain binary labels (\textit{COVID} vs.\ \textit{Non-COVID}), whereas the test set contains 1488 unlabeled scans. A major challenge of this benchmark is that the data are highly heterogeneous across sources, not only in acquisition and appearance, but also in class composition and source-specific sample balance.

\begin{table}[t]
\centering
\caption{Official dataset split.}
\label{tab:official_dataset}
\scriptsize
\setlength{\tabcolsep}{3.5pt}
\begin{tabular}{l l c c c c c}
\toprule
Split & Class & S0 & S1 & S2 & S3 & Total \\
\midrule
Train & COVID     & 175 & 175 & 39  & 175 & 564 \\
Train & Non-COVID & 165 & 165 & 165 & 165 & 660 \\
Val   & COVID     & 43  & 43  & 0   & 42  & 128 \\
Val   & Non-COVID & 45  & 45  & 45  & 45  & 180 \\
Test  & --        & --  & --  & --  & --  & 1488 \\
\bottomrule
\end{tabular}
\end{table}

However, we found that directly using the official split is suboptimal for reliable model development and source-aware evaluation. Therefore, we performed three practical corrections.

First, since the official validation split of Source 2 contains no positive COVID cases, source-specific validation on this source is ill-posed. To alleviate this issue, we additionally include the 39 positive Source 2 training scans in validation-oriented source analysis and model selection.

% Second, for Source 0 non-COVID training data, we found that the official \texttt{ct\_scan\_8} entry contains multiple folders, each corresponding to an individual CT sample. After folder-level inspection, the effective number of Source 0 non-COVID training scans is corrected from 165 to 231.

Second, for the Source 0 non-COVID training data, the official \texttt{ct\_scan\_8} entry was found to contain multiple folders, each representing an individual CT sample. Based on a folder-level inspection, the number of effective Source 0 non-COVID training scans was initially corrected from 165 to 231. Since \texttt{ct\_scan\_0} was confirmed to be absent, this number was subsequently adjusted to 230.

Third, since the challenge does not provide source labels for the test set, we estimate test-source identities using the Stage 3 source classifier. The raw predicted number of Source 0 test scans is 549, but because \texttt{ct\_scan\_492} also contains multiple CT samples and introduces ambiguity, we discard \texttt{ct\_scan\_492} and finally use 548 Source 0 test scans. The final predicted test-source distribution is 548/314/245/380 for Source 0/1/2/3, respectively, resulting in 1487 valid test scans used in our source-aware inference pipeline.

The corrected dataset statistics used in our experiments are summarized in Table~\ref{tab:revised_dataset}.

\begin{table}[t]
\centering
\caption{Revised train/val statistics used in our experiments.}
\label{tab:revised_dataset}
\scriptsize
\setlength{\tabcolsep}{3.5pt}
\begin{tabular}{l l c c c c c}
\toprule
Split & Class & S0 & S1 & S2 & S3 & Total \\
\midrule
Train & COVID     & 175 & 175 & 39  & 175 & 564 \\
Train & Non-COVID & 230 & 165 & 165 & 165 & 725 \\
Val   & COVID     & 43  & 43  & 39  & 42  & 167 \\
Val   & Non-COVID & 45  & 45  & 45  & 45  & 180 \\
\bottomrule
\end{tabular}
\end{table}
\subsection{Experiments}

We evaluate each stage of the proposed framework separately and report the best-performing configurations. Since the challenge emphasizes robustness across heterogeneous sources, we mainly focus on macro-F1, while also reporting accuracy (ACC) and area under the ROC curve (AUC) when available.

\subsubsection{Stage 1: Lung-Aware 3D Classification}

Stage~1 directly trains a 3D COVID classifier on canonicalized volumetric CT scans. We compare several input settings, including original CT volumes, lung-extracted CT volumes, and their combination, with or without additional rotation-based augmentation. Table~\ref{tab:stage1_results_full} summarizes the results.

% \begin{table*}[t]
% \centering
% \caption{Stage 1 results.}
% \label{tab:stage1_results_full}
% \scriptsize
% \setlength{\tabcolsep}{4pt}
% \begin{tabular}{l c c c c c c c}
% \toprule
% Setting & ACC & Macro-F1 & AUC & S0 & S1 & S2 & S3 \\
% \midrule
% Lung              & 0.9683 & 0.9682 & \textbf{0.9829} & 0.9886 & 0.9545 & 0.9523 & 0.9769 \\
% Lung + Rot        & 0.9568 & 0.9567 & 0.9803 & \textbf{1.0000} & 0.9204 & 0.9519 & 0.9540 \\
% Orig + Lung & \textbf{0.9712} & \textbf{0.9711} & 0.9791 & \textbf{1.0000} & 0.9545 & \textbf{0.9642} & 0.9654 \\
% Orig              & 0.9625 & 0.9625 & 0.9776 & 0.9773 & \textbf{0.9773} & 0.9152 & \textbf{0.9770} \\
% \bottomrule
% \end{tabular}
% \end{table*}

\begin{table*}[t]
\centering
\caption{Stage~1 results under different input configurations. \textit{Lung} denotes the lung-extracted CT dataset, \textit{Orig} denotes the original CT dataset, and \textit{Rot} denotes random rotation augmentation. S0–S3 denote source-wise F1 scores on the four sources.}
\label{tab:stage1_results_full}
\normalsize
\setlength{\tabcolsep}{10pt}
\renewcommand{\arraystretch}{1.2}
\begin{tabular}{lccccccc}
\toprule
Setting & ACC & Macro-F1 & AUC & S0 & S1 & S2 & S3 \\
\midrule
Lung        & 0.9683 & 0.9682 & \textbf{0.9829} & 0.9886 & 0.9545 & 0.9523 & 0.9769 \\
Lung + Rot  & 0.9568 & 0.9567 & 0.9803          & \textbf{1.0000} & 0.9204 & 0.9519 & 0.9540 \\
Orig + Lung & \textbf{0.9712} & \textbf{0.9711} & 0.9791 & \textbf{1.0000} & 0.9545 & \textbf{0.9642} & 0.9654 \\
Orig        & 0.9625 & 0.9625 & 0.9776          & 0.9773 & \textbf{0.9773} & 0.9152 & \textbf{0.9770} \\
\bottomrule
\end{tabular}
\end{table*}

As shown in Table~\ref{tab:stage1_results_full}, the best Stage~1 performance is obtained by combining \textit{Orig} and \textit{Lung}, which achieves the highest accuracy (\textbf{0.9712}) and macro-F1 (\textbf{0.9711}). This result indicates that the original scans and lung-focused scans capture complementary cues, and that their combination is more effective than either input alone. Notably, Stage~1 reaches a perfect score on Source~0, providing strong justification for using the Stage~1 expert alone on Source~0 in the final source-aware inference framework.

% As shown in Table~\ref{tab:stage1_results_full}, the best Stage~1 setting is obtained by combining original CT volumes and lung-extracted CT volumes with rotation-based augmentation, achieving a macro-F1 of \textbf{0.9711} and a validation accuracy of \textbf{0.9712}. Importantly, the Source 0 accuracy reaches \textbf{100\%}, which strongly supports our later decision to use the Stage~1 expert alone for Source 0 during source-aware inference.

\subsubsection{Stage 2a: Slice-Wise Representation and Probability Learning}

Stage~2a adapts a pretrained MedSigLIP encoder via slice-wise probability learning. We evaluate several input construction strategies, including contiguous random depth sampling from canonicalized $24 \times 448 \times 448$ scans, contiguous random depth sampling from $128 \times 256 \times 256$ scans, and a depth-random-sampling variant applied to the $24 \times 448 \times 448$ setting. The quantitative results are reported in Table~\ref{tab:stage2a_results}.

\begin{table}[t]
\centering
\caption{Stage~2a results using Orig + Lung inputs. ``CRS'' denotes contiguous random sampling and ``DRS'' denotes depth-random sampling.}
\label{tab:stage2a_results}
\normalsize
\setlength{\tabcolsep}{7pt}
\renewcommand{\arraystretch}{1.2}
\begin{tabular}{lccc}
\toprule
Setting & ACC & Macro-F1 & AUC \\
\midrule
CRS (24×448×448) & \textbf{0.9481} & \textbf{0.9450} & \textbf{0.9864} \\
CRS (128×256×256) & \textbf{0.9481} & 0.9425 & 0.9812 \\
DRS (24×448×448) & 0.9452 & 0.9440 & 0.9821 \\
\bottomrule
\end{tabular}
\end{table}

Overall, Stage~2a achieves strong performance, with the best macro-F1 reaching \textbf{0.9450} under contiguous random sampling and the highest AUC reaching \textbf{0.9864}. These results confirm that pretrained MedSigLIP can be effectively adapted to chest CT classification through slice-wise probability supervision.

\subsubsection{Stage 2b: Transformer-Based Inter-Slice Context Modeling}

Stage~2b further models inter-slice dependency on top of the pretrained MedSigLIP encoder from Stage~2a. We compare three settings: (1) training only the Transformer blocks and the new classification head, (2) jointly training the Transformer blocks, the classification head, and the last two MedSigLIP layers, and (3) directly flattening visual features followed by classification. The results are summarized in Table~\ref{tab:stage2b_results}.

% \begin{table*}[t]
% \centering
% \caption{Stage 2b results. ``Trans-only'' updates only the Transformer blocks and classifier; ``Trans + last2'' additionally updates the last two MedSigLIP layers; ``Flat + cls'' directly flattens visual features for classification.}
% \label{tab:stage2b_results}
% \scriptsize
% \setlength{\tabcolsep}{4pt}
% \begin{tabular}{l c c c c}
% \toprule
% Setting & Source-wise F1 (S0/S1/S2/S3) & Macro-F1 & ACC & AUC \\
% \midrule
% Trans-only   & 0.9524/0.9535/0.9231/0.9756 & 0.9511 & 0.9539 & 0.9837 \\
% Trans + last2 & 0.9524/0.9535/0.9250/0.9756 & 0.9516 & 0.9539 & \textbf{0.9854} \\
% Flat + cls   & \textbf{0.9647}/\textbf{0.9655}/\textbf{0.9268}/\textbf{0.9756} & \textbf{0.9582} & \textbf{0.9597} & 0.9847 \\
% \bottomrule
% \end{tabular}
% \end{table*}

\begin{table}[t]
\centering
\caption{Stage~2b results.}
\label{tab:stage2b_results}
\normalsize
\setlength{\tabcolsep}{7pt}
\renewcommand{\arraystretch}{1.2}
\begin{tabular}{lccc}
\toprule
Setting & Macro-F1 & ACC & AUC \\
\midrule
Trans-only    & 0.9511 & 0.9539 & 0.9837 \\
Trans + last2 & 0.9516 & 0.9539 & \textbf{0.9854} \\
Flat + cls    & \textbf{0.9582} & \textbf{0.9597} & 0.9847 \\
\bottomrule
\end{tabular}
\end{table}

Among the three settings, \textit{Flat + cls} delivers the best overall performance, achieving the highest macro-F1 (\textbf{0.9582}) and accuracy (\textbf{0.9597}). This result suggests that the pretrained MedSigLIP encoder already produces sufficiently discriminative slice representations, such that a lightweight classifier is enough to obtain strong classification performance. In contrast, \textit{Trans + last2}, which additionally trains the Transformer encoder, the last two MedSigLIP layers, and the newly introduced classification head, achieves the highest AUC (\textbf{0.9854}), indicating improved ranking ability.

\subsubsection{Stage 3: Source Classification}

Stage~3 predicts the source identity of each scan for source-aware expert routing. The source classifier is built on top of the Stage~1 3D backbone and evaluated on the validation split with the additional Source 2 positive samples described above. The final Stage 3 source classifier achieves an ACC of 0.9107 and an F1 score of 0.9114.

These results indicate that the source classifier can provide sufficiently reliable source predictions for downstream source-aware expert inference. Combined with the observation that Stage~1 achieves perfect performance on Source 0, this makes it possible to route Source 0 scans directly to the 3D expert, while using multi-expert fusion for Source 1/2/3.
\section{Conclusion}

In this work, we presented a three-stage source-aware multi-expert framework for robust multi-source COVID-19 CT classification. Instead of relying on a single unified classifier, our method explicitly addresses the key challenges of this benchmark, including strong inter-source heterogeneity, source imbalance, source-specific validation difficulty, and hidden test-source identities. The proposed framework integrates a lung-aware 3D expert for volumetric modeling, two MedSigLIP-based 2D experts for slice-wise adaptation and inter-slice context modeling, and a source classifier for source-aware routing and expert fusion.

The experimental results validate the effectiveness of this design. Stage 1 provides the strongest overall performance and achieves perfect accuracy on Source 0, which supports source-specific expert allocation. Stage 2 further contributes complementary discriminative cues through pretrained 2D visual modeling and inter-slice refinement. Stage 3 enables reliable source prediction and makes source-aware inference feasible on the unlabeled test set. By combining multiple experts and aggregating their outputs through hierarchical voting, our framework achieves robust and stable performance under challenging multi-source conditions.

Overall, our results suggest that multi-source COVID-19 CT detection should be addressed not only as a binary classification problem, but also as a source-aware domain-robust inference problem. We believe the proposed framework provides a practical and effective paradigm for building reliable medical imaging systems under real-world cross-institutional heterogeneity.

\paragraph{Acknowledgements.}
This work was supported by National Natural Science Foundation of China (No. 62576107), and the Shanghai Municipal Commission of Economy and Informatization, Corpus Construction for Large Language Models in Pediatric Respiratory Diseases (No.2024-GZL-RGZN-01013, and the Science and Technology Commission of Shanghai Municipality (No.24511104200), and 2025 National Major Science and Technology Project - Noncommunicable Chronic Diseases-National Science and Technology Major Project, Research on the Pathogenesis of Pancreatic Cancer and Novel Strategies for Precision Medicine (No.2025ZD0552303).
% This work was supported by National Natural Science Foundation of China (No. 62576107) Shanghai Municipal Commission of Economy and Informatization，Corpus Construction for Large Language Models in Pediatric Respiratory Diseases (No.2024-GZL-RGZN-01013)
% This work was supported (in part) by the Science and Technology Commission of Shanghai Municipality(No.24511104200)
% 2025 National Major Science and Technology Project —  Noncommunicable Chronic Diseases-National Science and Technology Major Project，Research on the Pathogenesis of Pancreatic Cancer and Novel Strategies for Precision Medicine (No.2025ZD0552303)

{
    \small
    \bibliographystyle{ieeenat_fullname}
    \bibliography{main}

@String(ICIP = {IEEE Int. Conf. Image Process.})

@String(ICIP  = {ICIP})

@inproceedings{li2024advancing,
  title={Advancing covid-19 detection in 3d ct scans},
  author={Li, Qingqiu and Yuan, Runtian and Hou, Junlin and Xu, Jilan and Zhang, Yuejie and Feng, Rui and Chen, Hao},
  booktitle={Proceedings of the IEEE/CVF Conference on Computer Vision and Pattern Recognition},
  pages={5149--5156},
  year={2024}
}

@inproceedings{yuan2024domain,
  title={Domain adaptation using pseudo labels for covid-19 detection},
  author={Yuan, Runtian and Li, Qingqiu and Hou, Junlin and Xu, Jilan and Zhang, Yuejie and Feng, Rui and Chen, Hao},
  booktitle={Proceedings of the IEEE/CVF Conference on Computer Vision and Pattern Recognition},
  pages={5141--5148},
  year={2024}
}

@inproceedings{li2025advancing,
  title={Advancing lung disease diagnosis in 3d ct scans},
  author={Li, Qingqiu and Yuan, Runtian and Hou, Junlin and Xu, Jilan and Zhang, Yuejie and Feng, Rui and Chen, Hao},
  booktitle={Proceedings of the IEEE/CVF International Conference on Computer Vision},
  pages={7377--7382},
  year={2025}
}

@inproceedings{kollias2025pharos, title={Pharos-afe-aimi: Multi-source \& fair disease diagnosis}, author={Kollias, Dimitrios and Arsenos, Anastasios and Kollias, Stefanos}, booktitle={Proceedings of the IEEE/CVF International Conference on Computer Vision}, pages={7265--7273}, year={2025}}

@inproceedings{kollias2024domain,
  title={Domain adaptation explainability \& fairness in ai for medical image analysis: Diagnosis of covid-19 based on 3-d chest ct-scans},
  author={Kollias, Dimitrios and Arsenos, Anastasios and Kollias, Stefanos},
  booktitle={Proceedings of the IEEE/CVF Conference on Computer Vision and Pattern Recognition},
  pages={4907--4914},
  year={2024}
}

@article{sellergren2025medgemma,
  title={Medgemma technical report},
  author={Sellergren, Andrew and Kazemzadeh, Sahar and Jaroensri, Tiam and Kiraly, Atilla and Traverse, Madeleine and Kohlberger, Timo and Xu, Shawn and Jamil, Fayaz and Hughes, C{\'\i}an and Lau, Charles and others},
  journal={arXiv preprint arXiv:2507.05201},
  year={2025}
}

@inproceedings{arsenos2022large,
  title={A large imaging database and novel deep neural architecture for covid-19 diagnosis},
  author={Arsenos, Anastasios and Kollias, Dimitrios and Kollias, Stefanos},
  booktitle={2022 IEEE 14th Image, Video, and Multidimensional Signal Processing Workshop (IVMSP)},
  pages={1--5},
  year={2022},
  organization={IEEE}
}

@inproceedings{arsenos2023data,
  title={Data-driven covid-19 detection through medical imaging},
  author={Arsenos, Anastasios and Davidhi, Andjoli and Kollias, Dimitrios and Prassopoulos, Panos and Kollias, Stefanos},
  booktitle={2023 IEEE International Conference on Acoustics, Speech, and Signal Processing Workshops (ICASSPW)},
  pages={1--5},
  year={2023},
  organization={IEEE}
}

@inproceedings{gerogiannis2024covid,
  title={Covid-19 computer-aided diagnosis through ai-assisted ct imaging analysis: Deploying a medical ai system},
  author={Gerogiannis, Demetris and Arsenos, Anastasios and Kollias, Dimitrios and Nikitopoulos, Dimitris and Kollias, Stefanos},
  booktitle={2024 IEEE International Symposium on Biomedical Imaging (ISBI)},
  pages={1--4},
  year={2024},
  organization={IEEE}
}

@article{kollias2018deep,
  title={Deep neural architectures for prediction in healthcare},
  author={Kollias, Dimitrios and Tagaris, Athanasios and Stafylopatis, Andreas and Kollias, Stefanos and Tagaris, Georgios},
  journal={Complex \& Intelligent Systems},
  volume={4},
  number={2},
  pages={119--131},
  year={2018},
  publisher={Springer}
}

@article{kollias2020deep,
  title={Deep transparent prediction through latent representation analysis},
  author={Kollias, Dimitrios and Bouas, N and Vlaxos, Y and Brillakis, V and Seferis, M and Kollia, Ilianna and Sukissian, Levon and Wingate, James and Kollias, S},
  journal={arXiv preprint arXiv:2009.07044},
  year={2020}
}

@inproceedings{kollias2020transparent,
  title={Transparent adaptation in deep medical image diagnosis},
  author={Kollias, Dimitris and Vlaxos, Y and Seferis, M and Kollia, Ilianna and Sukissian, Levon and Wingate, James and Kollias, S},
  booktitle={International Workshop on the Foundations of Trustworthy AI Integrating Learning, Optimization and Reasoning},
  pages={251--267},
  year={2020},
  organization={Springer}
}

@inproceedings{kollias2021mia,
  title={Mia-cov19d: Covid-19 detection through 3-d chest ct image analysis},
  author={Kollias, Dimitrios and Arsenos, Anastasios and Soukissian, Levon and Kollias, Stefanos},
  booktitle={Proceedings of the IEEE/CVF International Conference on Computer Vision},
  pages={537--544},
  year={2021}
}

@inproceedings{kollias2022ai,
  title={Ai-mia: Covid-19 detection and severity analysis through medical imaging},
  author={Kollias, Dimitrios and Arsenos, Anastasios and Kollias, Stefanos},
  booktitle={European conference on computer vision},
  pages={677--690},
  year={2022},
  organization={Springer}
}

@inproceedings{kollias2023ai,
  title={Ai-enabled analysis of 3-d ct scans for diagnosis of covid-19 \& its severity},
  author={Kollias, Dimitrios and Arsenos, Anastasios and Kollias, Stefanos},
  booktitle={2023 IEEE International Conference on Acoustics, Speech, and Signal Processing Workshops (ICASSPW)},
  pages={1--5},
  year={2023},
  organization={IEEE}
}

@article{kollias2023deep,
  title={A deep neural architecture for harmonizing 3-d input data analysis and decision making in medical imaging},
  author={Kollias, Dimitrios and Arsenos, Anastasios and Kollias, Stefanos},
  journal={Neurocomputing},
  volume={542},
  pages={126244},
  year={2023},
  publisher={Elsevier}
}

@article{kollias2024sam2clip2sam,
  title={Sam2clip2sam: Vision language model for segmentation of 3d ct scans for covid-19 detection},
  author={Kollias, Dimitrios and Arsenos, Anastasios and Wingate, James and Kollias, Stefanos},
  journal={arXiv preprint arXiv:2407.15728},
  year={2024}
}

@inproceedings{yuan2025multi,
  title={Multi-source covid-19 detection via variance risk extrapolation},
  author={Yuan, Runtian and Li, Qingqiu and Hou, Junlin and Xu, Jilan and Zhang, Yuejie and Feng, Rui and Chen, Hao},
  booktitle={Proceedings of the IEEE/CVF International Conference on Computer Vision},
  pages={7304--7311},
  year={2025}
}

@inproceedings{hou2021cmc,
  title={Cmc-cov19d: Contrastive mixup classification for covid-19 diagnosis},
  author={Hou, Junlin and Xu, Jilan and Feng, Rui and Zhang, Yuejie and Shan, Fei and Shi, Weiya},
  booktitle={Proceedings of the IEEE/CVF International Conference on Computer Vision},
  pages={454--461},
  year={2021}
}

@article{hou2021periphery,
  title={Periphery-aware COVID-19 diagnosis with contrastive representation enhancement},
  author={Hou, Junlin and Xu, Jilan and Jiang, Longquan and Du, Shanshan and Feng, Rui and Zhang, Yuejie and Shan, Fei and Xue, Xiangyang},
  journal={Pattern Recognition},
  volume={118},
  pages={108005},
  year={2021},
  publisher={Elsevier}
}

@inproceedings{hou2022cmc_v2,
  title={Cmc\_v2: Towards more accurate covid-19 detection with discriminative video priors},
  author={Hou, Junlin and Xu, Jilan and Zhang, Nan and Wang, Yi and Zhang, Yuejie and Zhang, Xiaobo and Feng, Rui},
  booktitle={European Conference on Computer Vision},
  pages={485--499},
  year={2022},
  organization={Springer}
}

@inproceedings{hou2022boosting,
  title={Boosting covid-19 severity detection with infection-aware contrastive mixup classification},
  author={Hou, Junlin and Xu, Jilan and Zhang, Nan and Zhang, Yuejie and Zhang, Xiaobo and Feng, Rui},
  booktitle={European Conference on Computer Vision},
  pages={537--551},
  year={2022},
  organization={Springer}
}

@article{gunraj2022covidnetct2,
  title={Covid-net ct-2: Enhanced deep neural networks for detection of covid-19 from chest ct images through bigger, more diverse learning},
  author={Gunraj, Hayden and Sabri, Ali and Koff, David and Wong, Alexander},
  journal={Frontiers in Medicine},
  volume={8},
  pages={729287},
  year={2022},
  publisher={Frontiers Media SA}
}

@inproceedings{krueger2021vrex,
  title={Out-of-distribution generalization via risk extrapolation (rex)},
  author={Krueger, David and Caballero, Ethan and Jacobsen, Joern-Henrik and Zhang, Amy and Binas, Jonathan and Zhang, Dinghuai and Le Priol, Remi and Courville, Aaron},
  booktitle={International conference on machine learning},
  pages={5815--5826},
  year={2021},
  organization={PMLR}
}

@article{zhang2017mixup,
  title={mixup: Beyond empirical risk minimization},
  author={Zhang, Hongyi and Cisse, Moustapha and Dauphin, Yann N and Lopez-Paz, David},
  journal={arXiv preprint arXiv:1710.09412},
  year={2017}
}

@inproceedings{wang2022medclip,
  title={Medclip: Contrastive learning from unpaired medical images and text},
  author={Wang, Zifeng and Wu, Zhenbang and Agarwal, Dinesh and Sun, Jimeng},
  booktitle={Proceedings of the 2022 Conference on Empirical Methods in Natural Language Processing},
  pages={3876--3887},
  year={2022}
}

@article{vaswani2017attention,
  title={Attention is all you need},
  author={Vaswani, Ashish and Shazeer, Noam and Parmar, Niki and Uszkoreit, Jakob and Jones, Llion and Gomez, Aidan N and Kaiser, {\L}ukasz and Polosukhin, Illia},
  journal={Advances in neural information processing systems},
  volume={30},
  year={2017}
}

@inproceedings{cao2021exploiting,
  title={Exploiting Deep Cross-Slice Features From CT Images For Multi-Class Pneumonia Classification},
  author={Cao, Jiawang and Jiang, Lulu and Hou, Junlin and Jiang, Longquan and Zhao, Ruiwei and Shi, Weiya and Shan, Fei and Feng, Rui},
  booktitle={2021 IEEE International Conference on Image Processing (ICIP)},
  pages={205--209},
  year={2021},
  organization={IEEE}
}
}

% WARNING: do not forget to delete the supplementary pages from your submission 
% \input{sec/X_suppl}

\end{document}